\documentclass{ws-p8-50x6-00}

\begin{document}

\title{Constructing Solutions of the Einstein Constraint Equations}

\author{James Isenberg}

\address{Department of Mathematics\\and\\Institute of Theoretical
Science\\University of Oregon\\ Eugene, OR  97403 USA\\  E-mail:
jim@newton.uoregon.edu}

\maketitle  

\abstracts{
The first step in the building of a spacetime solution of Einstein's
gravitational
field equations via the initial value formulation is finding a solution of the
Einstein constraint equations.  We recall the conformal method for constructing
solutions of the constraints and we recall what it tells us about the
parameterization of the space of such solutions.  One would like to know how to
construct solutions which model particular physical phenomena.  One useful step
towards this goal is learning how to glue together known solutions of the
constraint equations.  We discuss recent results concerning such gluing.}

\section{Introduction}
The constraint equations of Einstein's theory of the gravitational field are
familiar to most of those who work in general relativity:  If we choose a set
of appropriate local spacetime coordinates $(x^a, t)$ and if we write
Einstein's field equations for the gravitational metric field $g$ together
with a coupled matter source field\footnote{We presume that the matter source field is not
derivative-coupled, and that the coupled field equations constitute a
well-posed PDE system.} $\Psi$ in the tensorial form\footnote{We use
$MTW$\cite{MTW} conventions on signs of the curvature and on indices: Greek
letters run from 0 to 3 (spacetime) while Latin letters run from 1 to 3
(space).  Here $\kappa$ is a coupling constant.}
\begin{equation}
G_{\mu \nu} (g) = \kappa T_{\mu \nu} (g, \Psi) ,
\label{1}
\end{equation}
then the four equations
\begin{equation}
G_{t \nu} (g) = \kappa T_{t \nu} (g, \Psi) ,
\label{2}
\end{equation}
involve no time derivatives of higher than first order.  These are the Einstein
constraint equations.

What makes the constraint equations important in the study of the physics
of the
gravitational field is their role in the initial value formulation 
of  Einstein's equations.  We recall how the initial value formulation works:  Say we want to
construct a
spacetime solution $\{g, \Psi\}$ of the Einstein equations
(\ref{1}) on a manifold $M^4 = \Sigma^3 \times R $, where
$\Sigma^3$ is a smooth three-dimensional manifold.  The first step in doing
this
is to find a set of {\it initial data} consisting of\\
\begin{tabular}[h]{lcl}
$\gamma_{ab}$&-& a Riemannian metric on $\Sigma^3$\\
$K^{cd}$&-& a symmetric tensor field on $\Sigma^3$\\
$\psi$&-& space covariant pieces of the matter field $\Psi$ on
$\Sigma^3$\\
\end{tabular}\\
with $\{\gamma, K, \psi\}$ satisfying the constraint equations
(\ref{2}).  In terms of the data $\{\gamma, K,
\psi\}$ the constraint equations take the form
\begin{equation}
R + (tr K)^2 - K^{cd}K_{cd} = 2\kappa \rho(\gamma, \psi)
\label{3}
\end{equation}
\begin{equation}
\nabla_a K^a_b - \nabla _b (tr K) = \kappa J_b (\gamma, \psi)
\label{4}
\end{equation}
where $\nabla_a$ is the Levi-Civita\footnote{i.e., metric compatible and
torsion-free} covariant derivative corresponding to $\gamma, R$ is its scalar
curvature, $\rho$ is a scalar function of $\gamma$ and $\psi$, and $J_b$ is a
vector field function of $\gamma$ and $\psi$.\footnote{If, for example, the
``matter source field'' is Maxwell's electromagnetic field, then the spacetime
covariant field $\Psi$ is the spacetime one-form field $A_u$, the space
covariant
fields $\psi$ may be chosen to be the magnetic spatial one-form field $B_a$ and
the electric spatial vector field $E^a$, and one calculates $\rho = {1 \over 2}
(E^a E^b \gamma_{ab} + B^a B^b \gamma^{ab})$ and $J_c =
\eta_{cmn}E^mB^n$, where $\eta_{cmn}$ is the alternating symbol 
$\eta_{123} = 1$, etc.  In addition, the data must satisfy the Maxwell
constraint equations $\nabla^a B_a = 0$ and $\nabla_aE^a = 0$.}

Once we have obtained constraint-satisfying initial data $\{\gamma, K, \psi\}$,
we may evolve this data in time by choosing everywhere on $\Sigma^3$ a scalar
field $N$ (the ``lapse function'') and a vector field $M^a$ (the ``shift
vector'')
and then solving the Einstein evolution equations in the form

\begin{equation}
{d \over dt}\gamma_{ab} = - 2NK_{ab} +\pounds _M\gamma_{ab}
\label{5}
\end{equation}
and
\begin{equation}
{d \over dt} K^c_d = N(R^c_d +  trK K^c_d) - \nabla^c \nabla _d N + \kappa T^c_d + \pounds_M K^c_d
\label{6}
\end{equation}
where $T^c_d$ are the spatial components of the stress-energy tensor
corresponding to the matter field $\Psi$. We note that the time evolution of
the lapse and shift are not determined by the Einstein equations; they may
be chosen freely, and they must be chosen for all $t$ if one wishes to
evolve $\gamma$ and $K$ in time. The role of $N$ and $M$ is to
determine the coordinates relative to the evolving spacetime.

After $N(x,t)$ and $M^c(x,t)$ have been chosen and the evolved data
$\gamma_{ab}(x,t)$ and $K^{cd}(x,t)$ and $\psi(x,t)$ have been
determined\footnote{The coupled Einstein-matter source field equations
determine evolution equations for $\psi$.}, we may reconstruct the
spacetime metric (on the manifold $M^4=\Sigma^3 \times I $, where $I$ is the
interval for which the evolution can be carried out) via the formula
\begin{equation}
g(x, t) = - N^2dt^2 + \gamma_{ab} (dx^a + M^adt) (dx^b +
M^bdt).
\label{7}
\end{equation}
We similarly reconstruct $\Psi$.  It then follows (from the
Gauss-Codazzi-Mainardi 3+1 decomposition equations for the spacetime
curvature) that $g$ and $\Psi$ satisfy the Einstein equation (\ref{1}) on
$M^4$.

Does the initial value formulation work? This general question is best
considered by breaking it up into a number of more specific questions:
\begin{enumerate}
\item  Does every choice of initial data $\{\gamma, K, \psi\}$ which
satisfies the constraint equations evolve into a spacetime solution $\{g,
\Psi\}$?
\item  How do we find data $\{\gamma, K,\psi\}$ which satisfy the
constraint equations (\ref{3})-(\ref{4})?
\item  What is the space of solutions of the constraint equations?
\item  If we wish to use Einstein's equations to model the physical
behavior of a particular physical situation (e.g., a black hole collision), how
do we build that physical situation into the choice of initial data?
\item  What criteria should we use in making a choice of the lapse and
shift?
\item  What do we know about the long time behavior of solutions?
\item  Is the initial value formulation a practical way to construct and
study solutions of Einstein's equations?
\end{enumerate}

In this essay, we shall focus our attention on questions 2, 3, and 4, which
concern finding and understanding solutions to the constraint equations.
We start in Section 2 by discussing the conformal method for constructing
solutions to the constraints.  After recalling the basic steps of the
conformal method, we state the results which tell us the extent to which
the method is known to work.  Based on these results, we discuss (also in
Section 2) what is known about the function spaces which parameterize
the set of solutions.

Knowing how to construct constraint-satisfying initial data sets from
conformal data sets, and knowing how to parameterize the space of all
such initial data sets, does not tell us much about how to use solutions
constructed via the initial value formulation for studying physical questions.  With observational
data from LIGO and other gravitational radiation detectors expected to
start arriving within the next few years, it is crucial that we learn how to
do this.  Although progress to date on this front has been fairly
limited, recent results on the gluing  of initial data sets could be
useful.  We
discuss some of these gluing results and the ideas involved in Section 3.

Only if initial data can be evolved is it worthwhile to construct it.  We
 comment  briefly on evolution in Section 4, touching on questions 1,
5, 6, and 7.  We make some concluding remarks in Section 5.

\section{The Conformal Method and the Space of Solutions of the
Constraint Equations}

For the past thirty years, the conformal method has been the most
successful procedure for producing and studying solutions of the Einstein
constraint equations, both numerically and analytically.  The key idea of
this method, which was developed primarily by Lichnerowicz,
Choquet-Bruhat and York\cite{CBY} is to split the initial data $\{\gamma,
K,\psi\}$ into two parts:  The first part is to be chosen freely, while the
second part is to be determined by the constraints. As a consequence of the split, 
the constraint equations take the form of a determined PDE system of four equations for four unknowns
(equations (\ref{12})-(\ref{13}) below), rather than the form of an underdetermined system of four
equations for twelve unknowns (equations (\ref{3})-(\ref{4})).

To illustrate how this method works explicitly with matter source fields as
well as gravitational fields, let us consider the Einstein-Maxwell equations,
for which the initial data consists of $\{\gamma_{ab}, K^{cd}, B_a, E^a\}$
and for which the full set of constraint equations takes the form
(presuming no charges present)
\begin{equation}
R + (trK)^2 - K^{cd} K_{cd} = \kappa(B^aB_a + E^aE_a)
\label{8}
\end{equation}
\begin{equation}
\nabla_aK^a_b - \nabla_b (trK) = \kappa \eta_{bmn}B^m  E^n
\label{9}
\end{equation}
\begin{equation}
\nabla^aB_a  = 0
\label{10}
\end{equation}
\begin{equation}
\nabla_aE^a  = 0
\label{11}
\end{equation}For this system, the split of the data is as follows:
\begin{description}
\item[Free (``Conformal'') Data:]\
\begin{description}
\item $\lambda_{ab}$ - a Riemannian metric
(specified up to conformal factor)
\item$\sigma^{cd}$ - a divergence-free\footnote{In the conformal data,
the divergence-free condition is defined using the Levi-Civita covariant
derivative compatible with the conformal metric $\lambda_{cd}$.} and
trace-free symmetric tensor field\\
($\nabla _c\sigma^{cd} = 0$) \hspace{1 in} ($\lambda_{cd}c\sigma^{cd} =
0$)
\item $\tau$ - a scalar field
\item $\beta^a$ - a divergence-free vector field
\item $\varepsilon^a$ - a  divergence-free vector field
\end{description}
\item[Determined Data]\
\begin{description}
\item $\phi$ - a positive definite scalar field
\item  $W^a$ - a vector field
\end{description}
\item[Determining Equations]\
\begin{equation}
\nabla_a (LW)^a_b = {2 \over 3}\phi^6\nabla_b\tau + \kappa
\eta_{bmn}\beta^m\varepsilon^n
\label{12}
\end{equation}
\begin{eqnarray}
\nabla^2 \phi &=& {1 \over 8}R\phi - {1 \over 8}(\sigma^{ab} +
LW^{ab})(\sigma_{ab} + LW_{ab})\phi^{-7} \nonumber \\
&&-{\kappa \over 8} (\beta^a \beta_a + \varepsilon^a\varepsilon_a)\phi^{-3} + {1 \over
12}\tau^2 \phi^5
\label{13}
\end{eqnarray}
\end{description}
Here the covariant derivative, the Laplacian, the scalar curvature, and the
index manipulations all correspond to the metric $\lambda_{ab}$. Note
that $L$ is the conformal Killing operator, defined by
\begin{equation}
LW_{ab}:= \nabla_a W_b + \nabla_b W_a - {2 \over 3}\lambda_{ab}
\nabla_c W^c.
\label{14}
\end{equation}

The determining equations (\ref{12})-(\ref{13}) are to be solved for
$\phi$ and $W^a$.  If, for a given set of conformal data $\{\lambda_{ab},
\sigma^{cd}, \tau, \beta^a, \varepsilon^a\}$ one can indeed solve (\ref{12})-(\ref{13}),
then the reconstructed data
\begin{equation}
\gamma_{ab} = \phi^4\lambda_{ab}
\label{15}
\end{equation}
\begin{equation}
K^{cd} = \phi^{-10}(\sigma^{cd }+LW^{cd}) + {1 \over
3}\phi^{-4}\lambda^{cd}\tau
\label{16}
\end{equation}
\begin{equation}
B_a = \phi^{-2}\lambda_{ab}\beta^b
\label{17}
\end{equation}
\begin{equation}
E^a = \phi^{-6}\varepsilon^a
\label{18}
\end{equation}
constitute a solution of the constraint equations (\ref{8})-(\ref{11}).

Does the conformal method always work in the sense that, for any choice
of conformal data $\{\lambda_{ab}, \sigma^{cd}, \tau, \beta^a, \epsilon^a\}$ one
can always solve (\ref{12})-(\ref{13}) for $\{\phi , W^a\}$ and thereby
obtain a solution $\{\lambda_{ab}, K^{cd}, B_a, E^a\}$ for the constraints
(\ref{8})-(\ref{11})?  It is easy to see that this is not the case, as
illustrated
by the following simple example:

Let us choose the manifold $\Sigma^3$ to be the three-dimensional
sphere $S^3$, and let us choose the conformal data to consist of the round
sphere metric $\lambda = g_{(round)}$ with $R = 1$,  together with
$\sigma^{cd } = 0$, $\varepsilon^a = 0$, $\beta^a = 0$, and $\tau = 1$.
With $\tau$ constant and with $\eta_{bmn}\varepsilon^m \beta^n$ equal to
zero, the right hand side of equation (\ref{12}) vanishes.  Since the
operator $\nabla \cdot L$ is self-adjoint and injective (up to conformal Killing vector fields), 
it readily follows that the solutions
$W^a$ to (\ref{12}) for this choice of conformal data all satisfy the
condition $LW_{ab} = 0$ (which is the conformal Killing equation).  
Using this result, we find that the remaining
equation (\ref{13}) takes the form
\begin{equation}
\nabla^2 \phi = {1 \over 8} \phi  + {1 \over
12}  \phi^5.
\label{19}
\end{equation}
Since we seek solutions $\phi$ which are positive definite, for any such
solution the right hand side of equation (\ref{19}) is positive define.  But
the maximum principle tells us that there are no functions $\phi$ on a
compact manifold for which $\nabla^2 \phi > 0$ (or $<0$).  Hence, for this choice
of conformal data, there is no solution $\{\phi, W^a\}$ to equations
(\ref{12})-(\ref{13}), and therefore no corresponding solution to the
Einstein-Maxwell constraint equations.

This example shows that there are choices of conformal data which are not
mapped to solutions of the constraint equations by the conformal
method.  What happens generally?  This question is difficult to address in
any complete sense because there is a very wide variety of cases of
physical and mathematical interest, and each must be handled separately.
The various cases may be classified using the following criteria:

\begin{description}
\item[Manifold and Asymptotic Geometry]\
\begin{itemize}
\item Closed $\Sigma^3$
\item Asymptotically Euclidean
\item Asymptotically hyperbolic
\item Compact $\Sigma^3$ with boundary conditions
\end{itemize}
\item[Mean Curvature]\

\begin{itemize}
\item Maximal
\item Constant  non maximal (CMC)
\item Non Constant (non CMC)
\end{itemize}
\item[Fields]\
\begin{itemize}
\item Einstein vacuum
\item Einstein - standard matter source field\footnote{We label as
``standard matter source field'' any field theory which couples to
Einstein's theory without derivative coupling.  Included are
Einstein-Maxwell, Einstein-Yang-Mills, Einstein-Dirac,
Einstein-Klein-Gordon, Einstein-Cartan, and Einstein fluids.  For all these
theories, there are conformal splittings such that, if $\nabla\tau = 0$, the
constraint equations semi-decouple.  See Isenberg and Nester\cite {IN}. }
\end{itemize}
\item[Function Spaces]\
\begin{itemize}
\item $C^{\alpha}$ (analytic)
\item $C^{\infty}$ (smooth)
\item $C^{k + \beta}$ (H\"{o}lder)
\item $H^p_k$ (Sobolev)
\item $H^p_{k, \delta}$ (weighted Sobolev)
\end{itemize}
\end{description}

For a number of cases (labeled using these criteria), we have a fairly
complete understanding of which sets of conformal data map to solutions
and and which do not.  This is true for the following:
\begin{enumerate}
\item Closed $\Sigma^3$; CMC; Einstein vacuum or Einstein-standard
matter source field; $\lambda \in C^3$ and $\sigma$,
$\varepsilon$, $\beta \in H^p_2 (p > 3)$.
\item Asymptotically Euclidean; Maximal; Einstein vacuum or
Einstein-standard matter source field; Data in weighted Sobolev spaces.
\item Asymptotically hyperbolic: CMC; Einstein vacuum or
Einstein-standard matter source field; Data in weighted Sobolev spaces.
\end{enumerate}

For these three cases, one finds that all sets of conformal data which are
not readily disqualified by simple sign criteria or by the maximal principle
are mapped to solutions by the conformal method.  To explain the sense
in which this holds, let us focus for the moment on the first of these cases;
that with $\Sigma^3$ closed:

The Yamabe classification of metric conformal classes is a key tool for
determining which sets of conformal data on a closed manifold
$\Sigma^3$ are mapped to solutions of the constraints.  This classification
is based on the Yamabe Theorem\cite{Sch} which says that every metric
$\lambda _{ab}$ on a closed manifold $\Sigma^3$ is conformally related
to another which has constant scalar curvature either $+1$, $0$, or $-1$;
and moreover, the sign of $R$ for constant scalar curvature metrics
conformally related to a given $\lambda _{ab}$ is unique.  Thus the set of
all metrics (and metric conformal classes) on a closed $\Sigma^3$ is
partitioned into the three Yamabe classes ${\mathcal Y}^+(\Sigma^3)$,
${\mathcal Y}^0(\Sigma^3)$, and ${\mathcal Y}^-(\Sigma^3)$ depending
on this sign.

The Yamabe theorem and Yamabe classification are important for CMC
conformal data on a closed $\Sigma^3$ because solubility of
(\ref{12})-(\ref{13}) is invariant under the conformal transformation
$\{\lambda_{ab}, \sigma^{cd}, \tau, \beta^a, \varepsilon^a\} \rightarrow
\{\theta ^4\lambda_{ab}, \theta ^{-10}\sigma^{cd}, \tau,
\theta ^{-6}\beta^a, \theta ^{-6}\varepsilon^a\}$ for any positive $\theta$.  (This is {\it
not} true if $\nabla \tau \neq 0$).  Thus, to determine for which sets of
conformal data (\ref{12})-(\ref{13}) admit a solution, one may without loss of
generality restrict attention to conformal metrics with $R(\lambda) = -1, 0$, or
$+1$.

As a consequence of this conformal invariance and the resulting
simplification, one immediately sees from (\ref{13}) and the maximum
principle that the following categories of conformal data do not lead to
solutions\footnote {Here $(\sigma, \rho)\equiv 0$ means that $\sigma^{cd}$ and $\rho$ are identically zero everywhere on $\Sigma^3$, while $(\sigma, \rho)\not\equiv0$ means that there exists at least one point on $\Sigma^3$ at which $\sigma^{cd}$ or $\rho$ is not zero.} $\{{\cal Y}^+; (\sigma, \rho)\equiv 0; \tau = 0\}$,
$\{{\cal Y}^-; (\sigma, \rho)\equiv 0; \tau = 0\}$,
$\{{\cal Y}^+; (\sigma, \rho)\equiv 0; \tau \neq 0\}$,
$\{{\cal Y}^0; (\sigma, \rho)\equiv 0; \tau \neq 0\}$,
$\{{\cal Y}^0; (\sigma, \rho)\not\equiv 0; \tau = 0\}$,
$\{{\cal Y}^-; (\sigma, \rho)\not\equiv 0; \tau = 0\}$.
Much less evident is the fact that for {\it all other sets of conformal data
on closed $\Sigma^3$} (\ref{12})-(\ref{13}) can be solved, and we obtain a
solution of the constraints.  The proof of this result (with one case left out) is discussed  in the review paper of 
Choquet-Bruhat and York\cite{CBY}; the complete proof appears in the work of the author\cite{ICMC}.

Analogous results, with a few important modifications, hold for the other
two cases mentioned above.  Specifically, for the asymptotically Euclidean
case with $\tau = 0$, we first note that there is a Yamabe-type partition of
asymptotically Euclidean metrics into two classes $\cal Z^+$ and $\cal
Z^-$, with the $\cal Z^+$ conformal geometries admitting a conformal
transformation to metrics with $R=0$, and the $\cal Z^-$ conformal 
geometries not admitting such a
transformation.  It follows from work of Brill and Cantor\cite{BrCa} that
asymptotically Euclidean metrics with non-negative scalar curvature, as
well as those satisfying a certain integral condition, are contained in $\cal Z^+$.
The work of Cantor\cite{Ca} (See also\cite{CIY}) shows that a set of asymptotically
Euclidean conformal data $\{\lambda_{ab}, \sigma^{cd}, \tau, \beta^a,
\varepsilon^a\}$ (in appropriate function spaces) is mapped to a solution by the
conformal method if and only if $\lambda \in\cal Z^+$ (The
behavior of the other fields is irrelevant.).

For the asymptotically hyperbolic case (with $\tau \neq 0$), there are
important questions regarding the asymptotic behavior of solutions;
but ignoring those, the results are very simple:  For every set of asymptotically
hyperbolic conformal data, equations (\ref{12})-(\ref{13}) can be solved, and we
obtain a solution to the constraints\cite{AnChFr} \cite{AnCh}.

These results together tell us that if we are interested in constant mean
curvature (or maximal) initial data, the conformal method works, apart
from a few readily identified special cases.  This is useful for two reasons.
The first is that since equations of the form of the determining equations 
(\ref{12})-(\ref{13}) can be handled by modern computer algorithms (at least when they are 
semi-decoupled, which holds so long as we restrict to conformal data with  constant $\tau$),
 the conformal method 
is an effective way to produce CMC initial data sets for numerical relativity.
The second is that, as a consequence of uniqueness theorems
relating conformal data and solutions of the constraints\cite{ICMC}, together with
the results noted above, certain function spaces of conformal data provide
an effective parameterization of the set of solutions of the constraints\cite{IPRL}.  In
light of the uniqueness of CMC foliations of spacetime solutions of
Einstein's equations\cite{BrFl}, one may use these function spaces of conformal
data to parameterize the set of spacetime solutions as well (ignoring
those solutions which do not admit CMC or maximal foliations\cite{Br} \cite{Ba}).

These results are very useful, but they are limited in a number of ways, and so
there is a lot more that needs to be done.  First, one would like to know
which {\it non CMC} conformal data sets lead to solutions.  While there are
some results concerning this question, they all presume that
$|\nabla \tau|$ is small.  Indeed, for small $|\nabla \tau| $(with certain
restrictions on the zeros of $\tau$) our results show that solutions to the
coupled system (\ref{12})-(\ref{13}) always exist\cite{CBIM} \cite{IM} \cite{ICan} \cite{CIY} \cite{IP}. 
Unfortunately, the
iteration methods we have used to prove the results just cited do not
appear to be able to handle general $|\nabla \tau|$, so new ideas are likely
needed.

Next, we would like to be able to generate solutions to the constraint
equations on manifolds with boundaries.  The initial value-boundary value
problem has been studied for Einstein's theory in Friedrich's formulation\cite{FN}, 
but there has not been much mathematical work on it in the standard
$\{\gamma, K\}$ formulation discussed here.  In view of the recent
numerical work on spacetimes with excised black holes, a mathematical
treatment, in terms of the 
$\{\gamma, K\}$ formulation,  of solutions with boundaries would be useful.

A third issue regarding the construction of solutions of the constraint
equations which should be addressed further is the loosening of differentiability
requirements for solutions, including weak solutions.  The
motivation for doing this comes from the progress that has recently been
made by Klainerman and Rodnianski\cite{KR} and by Smith and Tataru\cite{ST} in
proving local well-posedness for the Einstein equations for data of lower
differentiability.  The goal (based on energy considerations) is
$H^2_{3/2}$; so far $H^2_{2 + \epsilon}$ has been obtained.  This leads
one to try to construct initial data sets in $H^2_{2 + \epsilon}$ or even
$H^2_{3/2}$.  The results to date\cite{ICMC} \cite{AnCh} \cite{CIY} produce initial data sets which
are considerably more restricted (i.e., smoother) than this, but work is
proceeding toward obtaining solutions of the constraint equations with
this low differentiability.

The question that remains is how to construct initial data sets which can
be used to model physical phenomena of interest.  We discuss this in the
next section.

\section{Physical Modeling and the Gluing of Solutions of the Constraint
Equations}

Say we want to study the gravitational radiation produced by the inspiral
collision of a neutron star and a black hole.  The first step of a numerical
modeling of this phenomenon is to find initial data $\{\gamma, K, \psi \}$
which constitutes a snapshot of the two objects and their ambient
spacetime at some moment prior to the collision.  It is important that the
initial data accurately represent the two objects and their surroundings in
a relatively quiet pre-collision state, without any extraneous unrelated radiation,
or the modeling will be of little use.

The electromagnetic analogue of this modeling problem is familiar and
easily handled, since the background space is fixed and independent of the
electromagnetic fields in Newtonian-Maxwell theory, and since the
representation of electromagnetic radiation is well understood in this
theory as well.  So, too, if we work with a post-Newtonian  approximation
to Einstein's theory, the choice of initial data for phenomena like the inspiral collision is
understood to some extent\cite{Da} \cite{BuDa} \cite{Alv}.

However, attempts to find physically accurate initial data sets for 
phenomenon such as this in terms of the full Einstein theory have been stymied,
despite many years of significant effort, both numerical and analytical.
The conformal method can of course be used to generate candidate sets of
data, and such sets are used in most numerical modeling studies.  The lack
of a priori control of the conformal factor and therefore of the ambient
space, together with the difficulty in representing and recognizing
gravitational radiation in the gravitational radiation in the conformal data
$\{\lambda,\sigma,\tau,\psi\}$ for massive relativistic objects renders
these sets somewhat suspect, however.

One approach that has been discussed to try to control the choice of initial data
is to work with sets of post-Newtonian initial data in the very early pre-radiation
stages of the collision and then try to evolve these into the later, more
relativistic stage. This approach has some promise, but can be very
expensive in numerical evolution time.

Another possible approach that could be tried relies on ``gluing." The idea
here is to start with well-understood initial data sets for separate portions
of the spacetime--say, one for the black hole and one for the neutron
star--and develop a procedure for joining these together into one set.
Mathematically, the idea is to join two or more solutions of the constraint
equations into a single solution in such a way that, away from the joining
region, the data is largely unchanged. Physically, the challenge is to
understand and control the physical effects of the joining region.

There has been substantial progress during the past couple of years on the
mathematical aspects of gluing solutions of the Einstein constraint
equations. Mazzeo, Pollack, and the author (IMP) have proven theorems
which prescribe a procedure for gluing connected sums of sets of initial
data\cite{IMP}, while Corvino and Schoen\cite{Cor} \cite{CorSch} have shown that rather general
asymptotically flat initial data sets can be glued to exact Schwarzschild or
exact Kerr exteriors outside a transition region. We first discuss the IMP
work, and then briefly describe the results of Corvino and Schoen.

\subsection{Connected Sum Gluing}

We start with a pair of constant mean curvature solutions $\{\Sigma^3_1,
\gamma_1, K_1\}$ and $\{\Sigma^3_2, \gamma_2, K_2\}$ of the vacuum
constraint equations (3)-(4). (We presume for now that there are no matter source fields $\psi$, 
and so $\rho=0$ and $J=0$).  Each of these solutions may be asymptotically
Euclidean, asymptotically hyperbolic, or one may have either or both of
$\Sigma^3_1$ and $\Sigma^3_2$ closed. The two solutions need not have
matching asymptotic properties, but they do need to have the same
constant mean curvature $\tau$.

If we now pick a pair of points $p_1 \in \Sigma ^3_1$,
$p_2 \in \Sigma^3_2$ on each manifold, then the gluing
procedure produces a one-parameter family of solutions
$\{\tilde{\Sigma}^3, \tilde{\gamma}_s, \tilde{K}_s \}$ with the following
properties:  (i)  $\tilde{\Sigma}^3 = \Sigma^3_1 \# \Sigma^3_2$,  with
$\Sigma^3_1$ joined to $\Sigma^3_2$ by a ``neck''
($S^2 \times R$) connecting a neighborhood of $p_1$ to a
neighborhood of
$p_2$ (both now excised); (ii) the data $\{\tilde{\gamma}_s, \tilde{K}_s \}$ can be made to 
be  arbitrarily close to the original data
$\{\gamma_1, K_1\}$ and $\{\gamma_2, K_2\}$ on appropriate regions
of  $\tilde{\Sigma}^3$ away from the neck by choosing $s$ sufficiently large; (iii) $tr \tilde{K} = \tau$; and
(iv) the geometry on the neck (i.e., lengths, curvature) can degenerate for
$s \rightarrow \infty$, but its behavior is controlled by exponential
bounds in $s$.

This gluing procedure works for quite general classes of data. The only restrictions
 on $\{\Sigma^3_1, \gamma_1, K_1\}$ and $\{\Sigma^3_2, \gamma_2, K_2\} $ besides the
CMC matching condition $trK_1=trK_2$ and certain regularity restrictions (stated in terms
of weighted H\"{o}lder spaces \cite{IMP}) are that if either of the sets of data has $\Sigma^3$ closed, 
that set of data must have $K^{cd}$ not vanishing identically, and must be free of nontrivial
conformal Killing vector fields\footnote {Actually, a Killing vector field can be present, so long 
as it does not vanish at the chosen points $p_1$ or $p_2$.}. Neither of these latter conditions
are needed for asymptotically Euclidean or asymptotically hyperbolic data, and we note that 
if a set of data $\{\Sigma^3, \gamma, K \}$ fails to satisfy either condition, it will generally satisfy
 them both after a small perturbation is made. 

The procedure allows a number of interesting sets of initial data to be produced. One can,
for example, add a sequence of small black holes to any asymptotically flat spacetime
$\{M^4, g\}$ by gluing a sequence of copies of Minkowski data to asymptotically Euclidean data for 
$\{M^4, g\}$ and then evolving the glued data. These spacetimes (somewhat
reminiscent  of the Misner multi-black hole solutions\cite{Mis}, but far more general) may be constructed 
with the interiors either independent
or connected. Another set of spacetimes one obtains via this gluing procedure 
allows one to study the question of black holes in cosmological spacetimes: One can glue 
asymptotically Euclidean or asymptotically hyperbolic data to cosmological data on closed $\Sigma^3$
and then see if black hole-like physics develops\footnote{Note that there is no accepted definition
of a black hole in a spacetime without some sort ''asymptotic infinity" structure.}. This gluing procedure 
also permits us to add one or more wormholes to most solutions 
of Einstein's equations: If one chooses the pair of points $p_1$ and $p_2$ on a single connected 
initial data set $\{\Sigma^3, \gamma, K \}$ and carries out the gluing procedure, one obtains initial 
data much like $\{\Sigma^3, \gamma, K \}$ except with an $S^2\times R$ wormhole glued on.

To what extent can one carry out these gluings in practice, say, numerically? To answer this question, 
we shall now describe the basic steps of the connected sum gluing procedure:

{\bf Step 1}: One starts by performing a conformal transformation of the data $\gamma_{ab} \rightarrow
\psi^{-4} \gamma_{ab}, \sigma^{cd} \rightarrow \psi^{10} \sigma^{cd}$, (Here $\sigma^{cd}$ is the
trace-free part of $K^{cd}$.) which is trivial ($\psi=1$) away from  a neighborhood of each of the points
$p_1$ and $p_2$, and which is singular at each of the points (with a specified type of singular behavior). 
The effect of this conformal 
blowup is to replace a ball surrounding $p_1$ with an $S^2\times R$ infinite length half-cylinder which is 
asymptotically a standard round half-cylinder, and to do the same in a neighborhood of $p_2$. 
Note that there are standard explicit formulas one can choose for $\psi$ to carry out this first step.

{\bf Step 2}: Each of the newly added half cylinders extends out infinitely far. If one cuts 
each at a distance $s/2$ (measured using the local conformally transformed metric) 
out from the neighborhoods of $p_1$ and $p_2$, and identifies the two cut half cylinders
on the joining sphere, then one has the manifold 
$\tilde{\Sigma}^3 = \Sigma^3_1 \# \Sigma^3_2$, with its connecting $s$-parametrized ''neck".
Then, using a pair of  cut-off 
functions $\chi_1$ and $\chi_2$, one may patch together the metric  
$\hat{\gamma}_s=\gamma_1\chi_1+\gamma_2\chi_2$ and the traceless apart 
of the second fundamental form $\hat{\sigma} _s=\sigma_1\chi_1+\sigma_2\chi_2$. Both $\hat{\gamma}_s$ 
and $\hat{\sigma}_s$ are well-defined on $\tilde{\Sigma}^3$, and both depend on 
the gluing distance parameter $s$. One also glues together the conformal 
function $\hat{\psi}_s=\psi_1\xi_1+\psi_2\xi_2$, using a different set of  cut-off functions. (For details, 
see  IMP\cite{IMP}). This step, like the first, can be done using standard explicit formulas for $\chi_1, \chi_2, 
\xi_1$, and $\xi_2$.

{\bf Step 3}: A careful choice of the cut-off functions results in $\hat{\sigma}_s$ being everywhere traceless 
with respect to $\hat{\gamma}_s$.  The conformal transformation $\sigma^{cd} \rightarrow \psi^{10} \sigma^{cd}$
guarantees that away from the neck, $\hat{\sigma}_s$ is divergence-free with respect to $\hat{\gamma}_s$. 
However, in the neck, $\hat{\nabla}_c \hat{\sigma}_s^{cd}\neq0$. So one replaces $\hat{\sigma}_s^{cd}$ by 
$\hat{\mu}_s^{cd},$ which is obtained by solving
\begin{equation}
\hat{\nabla}_c(\hat{L}Y_s)^{cd}=\hat{\nabla}_c\hat{\sigma}_s^{cd}
\end{equation} for the vector field $Y_s^c$, and setting 
\label{20}
\begin{equation}
\hat{\mu}_s^{cd}=\hat{\sigma}_s^{cd}-(\hat{L}Y_s)^{cd}.
\label{21}
\end{equation} The tensor field $\hat{\mu}_s^{cd} $ is divergence-free everywhere on $\tilde{\Sigma}^3$
 with respect to $\hat{\gamma}_s$. This step is straightforward to carry out in practice. However, to be able to proceed further with the gluing procedure, one needs to know not only that the divergence-free field
$\hat{\mu}_s^{cd} $ exists, but also that  $\parallel \hat{\mu}_s-\hat{\sigma}_s\parallel_{C^{k+\beta}} $
is very small for sufficiently large $s$. It is shown in IMP that this estimate always holds.

{\bf Step 4}: $\{\hat{\gamma}_s, \hat{\mu}_s, \tau\}$ is a standard set of conformal data on $\tilde{\Sigma}^3$. 
So one may set up and attempt to solve 
\begin{equation} 
\hat{\nabla}^2 \phi_s=\frac{1}{8}\hat{R}_s\phi_s-\frac{1}{8}\hat{\mu}_{scd}\hat{\mu}_s^{cd} \phi_s^{-7}+\frac{1}{12}\tau^2
\phi_s^5
\label{22}
\end{equation} for $\phi_s$. Although the Yamabe class of the metric $\hat{\gamma}_s$ constructed in Step 2 
is not evident, the 
estimates obtained in Step 3 allow one to show that (\ref{22}) has a unique solution. Further, with a 
considerable amount of work, one shows that $\parallel \phi_s-\psi_s\parallel_{C^{k+\beta}}$ is 
very small for sufficiently large $s$. As noted above, once one knows that a solution to (\ref{22}) exists,
one can readily obtain that solution numerically.

{\bf Step 5}: Using formulas analogous to (\ref{15})-(\ref{16}), one constructs $\tilde{\gamma}_s$ and 
$\tilde{K}_s$ from $\phi_s$ and $\{\hat{\gamma}_s, \hat{\mu}_s, \tau\}$. One immediately verifies 
that for each value of the parameter $s$, $\{ \tilde{\Sigma}^3, \tilde {\gamma}_s, \tilde {K}_s\}$ solves the vacuum constraint equations. As a consequence of the estimates of Steps 3 and 4, one verifies that away from the neck, $\hat{\gamma}_s$ and 
$\hat{K}_s $ approach the original data for sufficently large $s$.

We see from the description of these five steps that in practice, this connected sum gluing
procedure involves making a few choices of standard cut off functions, plus solving familiar
elliptic equations. It should be a useful procedure for studying solutions of Einstein's equations numerically. 

We note that,  to date, the IMP connected sum gluing procedure has been proven to work for 
the vacuum Einstein case only. It is expected that it can be implemented for Einstein-Maxwell, 
Einstein-Yang-Mills, and the other Einstein-standard matter source fields as well; work is under way to show this. It would be nice to also extend the procedure to non CMC data, but this is likely to be difficult.

\subsection{Exterior Schwarzschild and Kerr Gluing}

The starting point for the exterior Schwarzschild gluing procedure  is any asymptotically
 Euclidean, time symmetric ($K^{cd}=0$) initial data set $\{ \Sigma^3, \gamma\}$. 
One chooses a compact region $\Lambda^3$ in $\Sigma^3$. One then shows \cite{Cor} that 
there is an asymptotically Euclidean time symmetric solution $\{ \Sigma^3, \tilde{\gamma}\}$ such that 
(i) inside a ball $B_r$ which contains $\Lambda^3$ as a subset,  $\tilde{\gamma}|_{B_r}=\gamma$; and 
(ii) outside the ball $B_{2r}$, we have $\tilde{\gamma} |_{\Sigma^3\setminus B_{2r}} = \gamma_{Sch(m)}$, 
where  $\gamma_{Sch(m)}$ is the spatial Schwarzschild metric for some positive mass $m$.          

We note three important features of this type of gluing: First, when the gluing is completed, 
the region inside $B_r$ as well as the exterior region outside $B_{2r}$ are {\it unchanged}. 
This is not the case for the connected sum gluing, in which the original solutions are 
slightly changed (away from the neck) by the gluing procedure. Second, the analysis used to 
prove this result does not rely in any way on the conformal method. That is, one does not construct
the solution $\{\Sigma^3, \tilde{\gamma}\}$ by first specifying a conformal metric everywhere and 
then solving for the conformal factor. Indeed, one can carry out exterior Schwarzschild gluing only if one 
works with the constraints as an  underdetermined 
system to be solved for $\tilde{\gamma}$, rather than as a determined system to be solved for $\phi$. 
Third, the proof that this gluing works  does not readily translate into a step by step procedure that one 
can carry out numerically. The proof guarantees existence, but does not prescribe a procedure for constructing
$\{\Sigma^3, \tilde{\gamma}\}$

The result is quite surprising and remarkable. It says that for any compact portion of a time symmetric 
asymptotically Euclidean solution, one can arrange the gravitational fields in an annular region around
it so that outside this region, the gravitational field is exactly Schwarzschild. Note that 
the analogous result does not hold true for Newtonian gravity. Note also that, 
combining this result with the work of Friedrich\cite{Fri}, one may be able to produce 
a very large family of solutions of Einstein's equations with a complete ${\cal I}$ (null asymptotic infinity).

One large but unsurprising restriction on the type of data which can be glued to an 
exterior Schwarzschild spatial slice is that the data have $K^{cd}=0$. More recent 
results of Corvino and Schoen \cite{} allow one to remove this restriction provided 
that one replaces the Schwarzschild exterior by a slice of the more general Kerr (rotating
black hole) They do impose restrictions on the data $\{ \Sigma^3, \gamma, K\}$
 which can be patched to an exterior Kerr slice: They require certain decay behavior in 
$\gamma$ and $K$ which are essentially just enough so that the momentum and 
angular momentum integrals at infinity are well-defined. These restrictions appear to be minor, 
in view of the result. Again, one consequence of these results might be an even larger family 
of solutions with complete ${\cal I}$.

\section{Some Comments on Evolution}

Of the seven questions raised in the introduction, we have focussed in this
essay on those three -- 2,3, and 4 -- which deal with the constraint equations.
We shall now briefly comment on questions 1,5, and 6, which mostly
concern evolving sets of initial data into spacetime solutions.

The first question addresses the issue of local existence and well-posedness
of the initial value formulation of the Einstein equations. To a large extent,
this issue was settled fifty years ago by Choquet-Bruhat\cite{Br} who proved
(using harmonic coordinates) that for smooth initial data ($C^l$, with large
$l$) the Einstein system is well-posed. While this result is sufficient for
most purposes, there have been strong attempts in recent years to prove
well-posedness for more general classes of data. The goal is to prove it for
data in the Sobolev space $H^2_{\frac{3}{2}}$. The recent work of Klainerman and
Rodnianski\cite{KR}, and of Smith and Tataru\cite{ST}achieves it for
$H^2_{2+\epsilon} (\epsilon>0)$ data.

Well-posedness theorems show (among other things) that data in the
specified function space can always be evolved to a spacetime solution for a
sufficiently small time duration into the future and into the past. They say
little about how long a spacetime solution will last, or what its behavior will
be far into the future and far into the past. While these issues are very
difficult and far from resolved, there has been substantial progress on them
in recent years.  We note in particular the work of Christodoulou and
Klainerman\cite{ChrK}, which verifies the stability of Minkowski spacetime (and thereby proves that 
there is a nontrivial family of asymptotically flat spacetimes which extend for infinite 
proper time into the future and into the past);
that of Andersson and Moncrief\cite{AndM} which does the same for the stability of
the expanding $k=-1$ vacuum Friedman-Robertson-Walker 
cosmological spacetimes; and the work of Christodoulou\cite{Chr}
which verifies weak cosmic censorship\footnote{The weak cosmic censorship
conjecture says that in asymptotically Euclidean spacetime solutions, the
singularities which develop during gravitational collapse are generically
shielded from the view of observers at infinity by the development of an
event horizon. } for spherical collapsing Einstein-scalar field spacetimes.
There has also been a collection of works by Berger, Chrusciel,
Garfinkle, Kichenassamy, Moncrief, Rendall, Ringstrom, Wainwright,
Weaver, and the author which provide increasing evidence for the
presence of asymptotically velocity dominated behavior and mixmaster
behavior in cosmological spacetimes, and further tends to support the
validity of the strong cosmic censorship conjecture\footnote{The strong 
cosmic censorship conjecture\cite{Pen} says that in generic
solutions developed from Cauchy data on a compact Cauchy surface, a
Cauchy horizon (with its attendant causality difficulties) does not form.}
in these spacetimes.
(See, for example, the review papers by Berger\cite{Ber} and Rendall\cite{Ren}.)

The construction or study of spacetime solutions of Einstein's equations via
the initial value formulation requires one to make a choice of spacetime
foliation (i.e., a specification of the $t=$ constant hypersurfaces which fill
the spacetime) and of spacetime threading (i.e., a specification of the
$x^a =$ constant observer world lines which are everywhere transverse to
the foliation, and fill the spacetime). The choice of foliation is controlled
infinitesmally by the lapse function $N$, and the choice of threading is
similarly controlled by the shift vector $M^a$.

In constructing a spacetime from specified initial data, one may always
make the simple choice $N=1$ and $M^a =0$ (``Gaussian normal
coordinates'').  However, this choice generally leads to a non physical and
premature breakdown in the evolution (a coordinate singularity).
Consequently, the maximal development\cite{CBG}  of the initial data may not be
obtained. So, one criterion for a good choice of the lapse and shift is that
the choice effectively avoids such coordinate singularities. Also important
is that the choice be relatively easy to implement in practice, and that it not
obscure the gravitational physics of the spacetime (e.g., by simulating
gravitational radiation which is not physically present).

Constant mean curvature or maximal slicing, coupled with some sort of
coordinate shear minimizer, is often cited as a choice of foliation and
threading which avoids coordinate singularities and clarifies the physics.
However, implementing this choice requires that one solve a set of elliptic
equations on each time slice. This is a large expense in computer time for
numerical constructions, and it precludes explicit forms for the lapse and
shift in analytical studies. For special families of solutions such as the
Gowdy spacetimes  or the $U(1)$ Symmetric solutions , there are
certain choices of lapse and shift picked out by the geometry (areal for
Gowdy\cite{Gow}, harmonic time for $U(1)$ Symmetric\cite{Mon}); but these are special cases.
More generally, the choice of lapse and shift remains a difficult issue.

\section{Concluding Remarks}

We have discussed a number of the challenges  that one encounters in
constructing and studying spacetimes via the initial value formulation.
These occur both in finding sets of initial data which satisfy the Einstein
constraint equation, and in evolving these sets of data. Some are fairly difficult.
However, paraphrasing Winston Churchill's description of democracy, one finds
that the initial value formulation is the most impractical way to work with
solutions of Einstein's equations, except all of those other ways which 
have been tried from time to time\cite{WC}. The fact
that most numerical studies of solutions are carried out using the initial
value formulation attests to the relative practicality of this approach.

\section{Acknowledgments}

I thank the Scientific Committee of GR16 for inviting me to speak, and I
thank the local Organizing Committee and the local South African hosts for
running a very fine conference. Portions of the work discussed have been
supported by NSF grant PHY 0099373.

\end{document}